\documentclass[12pt,a4paper]{revtex4}
\usepackage{amsmath,amssymb,graphics,epsfig,subfigure}
\usepackage{color}
\usepackage[colorlinks,linkcolor=red,anchorcolor=red,citecolor=green]{hyperref}
\usepackage{geometry}
\makeatletter

\newcommand{\Rmnum}[1]{\expandafter\@slowromancap\romannumeral #1@}
\makeatother

\begin{document}

\thispagestyle{empty}

\begin{center}

\title{Diagnosis inspired by the thermodynamic geometry for different thermodynamic schemes of the charged BTZ black hole}

%\date{\today}
\author{Zhen-Ming Xu$^{1,2,3,4,}$\footnote{E-mail: xuzhenm@nwu.edu.cn}, Bin Wu$^{1,2,3,4,}$\footnote{E-mail: binwu@nwu.edu.cn (Corresponding author)} and Wen-Li Yang$^{1,2,3,4,}$\footnote{E-mail: wlyang@nwu.edu.cn}}

\affiliation{ $^{1}$Institute of Modern Physics, Northwest University, Xi'an 710127, China\\
$^{2}$School of Physics, Northwest University, Xi'an 710127, China\\
$^{3}$Shaanxi Key Laboratory for Theoretical Physics Frontiers, Xi'an 710127, China\\
$^{4}$Peng Huanwu Center for Fundamental Theory, Xi'an 710127, China}

\begin{abstract}
Due to the asymptotic structure of the black hole solution, there are two different thermodynamic schemes for the charged Banados-Teitelboim-Zanelli (BTZ) black hole. In one scheme, the charged BTZ black hole is super-entropic, while in the other, it is not (the reverse isoperimetric inequality is saturated). In this paper, we investigate the thermodynamic curvature of the charged BTZ black hole in different coordinate spaces. We find that in both schemes, the thermodynamic curvature is always positive, which may be related to the information of repulsive interaction between black hole molecules for the charged BTZ black hole if we accept an empirical relationship between the thermodynamic curvature and interaction of a system. More importantly, we provide a diagnosis for the discrimination of the two schemes from the point of view of the thermodynamics geometry. For the charged BTZ black hole, when the reverse isoperimetric inequality is saturated, the thermodynamic curvature of an extreme black hole tends to be infinity, while when the reverse isoperimetric inequality is violated, the thermodynamic curvature of the extreme black hole goes to a finite value.
\end{abstract}

\maketitle
\end{center}

\section{Introduction}
At present, black hole physics is generally considered as one of the best and effective ways to explore quantum gravity. Especially with the  pioneering discovery of Hawking and Bekenstein about the temperature and entropy in the black hole \cite{Hawking1975,Hawking1983,Bekenstein1973,Bardeen1973}, the general relativity, quantum mechanics and statistical physics are closely linked together to make it possible for us to glimpse the tip of the iceberg of quantum gravity. Thermodynamics theory, which has been rested on general principles spanning a wide range of pure fluid physical systems, is now applied to black hole systems extensively and successfully \cite{Wald2001}. One of the most prominent is the introduction of the extended phase space \cite{Kastor2009,Dolan2011a,Dolan2011b}, which makes the charged AdS black hole and the van der Waals fluid have a close relationship \cite{Kubiznak2012,Spallucci2013}. Black holes exhibit abundant phase transitions and critical behaviors in such an extended phase space \cite{Carlip2014,Kubiznak2017,Altamirano2014}.

Recently, the theory of the thermodynamics geometry \cite{Ruppeiner1995,Ruppeiner2007,Ruppeiner2014,Ruppeiner2010} is widely applied to the thermodynamics system of black holes, which provides a new and more traceable perspective for studying the micro-mechanism of black holes from the axioms of thermodynamics phenomenologically. This new scheme is mainly to use the Hessian matrix structure to represent the thermodynamic fluctuation theory \cite{Ruppeiner1995}. Thermodynamic curvature is the most important physical quantity in the theory. Taking advantage of the new scheme, the primary microscopic information of the Banados-Teitelboim-Zanelli (BTZ) black hole, Schwarzschild (-AdS) black hole, Reissner-Nordstr\"{o}m (-AdS) black hole, Gauss-Bonnet (-AdS) black hole, higher dimensional black holes and other black holes are explored \cite{Ruppeiner2008,Dolan2015,Wei2015,Wei2019a,Wei2019b,Wei2019c,Miao2018a,Miao2018b,Miao2019a,Miao2019b,Aman2003,Mirza2007,Dehyadegari2017,Cai1999,Zhang2015a,
Zhang2015b,Liu2010,Xu2019a,Xu2019b,Niu2012,Wang2019,Ghosh2019,Bhattacharya2017,Chen2019,Guo2019,Mansoori2014,Mansoori2015,Mansoori2016}.

In the early researches \cite{Cai1999,Sarkar2006,Quevedo2009,Wei2009,Akbar2011,Mohammadzadeh2018} on the thermodynamic geometry of BTZ black holes, the negative cosmological constant was regarded as a fixed parameter or fluctuation parameter. Such a system has no pressure and volume terms, that is, no extended phase space. Nowadays the notion of the extended phase space has been widely recognized and a good thermodynamic system must have pressure and volume terms. Therefore, it is necessary to discuss the thermodynamic geometry of the BTZ black hole again in the extended phase space.

We note that the study~\cite{Frassino2015} pointed out that due to the asymptotic structure of the BTZ black hole solution, it renders computation of the mass of charged BTZ black hole more problematic. They provided two schemes for analyzing the thermodynamic behavior of the charged BTZ black hole. In one case, the charged BTZ black hole is super-entropic \cite{Johnson2019a,Johnson2019b}, that is the reverse isoperimetric inequality is violated. While in the other, the reverse isoperimetric inequality is satisfied. Therefore, it is natural to ask how the thermodynamic geometry of the charged BTZ black hole is different or the same under different thermodynamic schemes. At present, this research content is still lacking. Hence in this paper, we shall explore the thermodynamic geometry of $(2+1)$-dimensional charged BTZ black hole in different coordinate spaces under two thermodynamic schemes in the framework of the extended phase space. Meanwhile, we provide a diagnosis for the discrimination of the two schemes from the point of view of thermodynamics geometry. This is the main motivation of our work. We find that in both schemes, the thermodynamic curvature is always positive. For the charged BTZ black hole, when the reverse isoperimetric inequality is saturated, the thermodynamic curvature of an extreme black hole tends to be infinity, while when the reverse isoperimetric inequality is violated, the thermodynamic curvature of the extreme black hole goes to a finite value.

Coincidentally, in the extended phase space, the study \cite{Ghosh2020} looked at the behavior of the thermodynamic curvature of the charged and rotating BTZ black hole with respect to entropy and thermodynamic volume respectively. Meanwhile, the authors of Ref.~\cite{Ghosh2020} extended the study to the case of the exotic BTZ black hole. We notice that in \cite{Ghosh2020} the thermodynamic volume and entropy are independent, i.e., this situation corresponds to our Case \Rmnum{1} (see the following for details). In our present paper, we discuss the thermodynamic curvature of the charged BTZ black hole in different coordinate spaces under two thermodynamic schemes (Case \Rmnum{1} and Case \Rmnum{2}) and provide a diagnosis for the discrimination of the two schemes from the point of view of the thermodynamics geometry. For the charged BTZ black hole without any angular momentum, our results are consistent with those of Ref.~\cite{Ghosh2020}, where a positive thermodynamic curvature has been obtained.

The paper is organized as follows. In section \ref{sec2}, we briefly review the Ruppeiner thermodynamic geometry. In section \ref{sec3}, we give two different thermodynamic schemes of the charged BTZ black hole. In section \ref{sec4}, we analyze the behavior of the thermodynamic curvature for the charged BTZ black hole under different thermodynamic schemes. Finally, we devote to drawing our conclusion and making further discussion in section \ref{sec5}. Throughout this paper, we adopt the units $\hbar=c=k_{_{B}}=G=1$.

\section{Ruppeiner thermodynamic geometry}\label{sec2}
Completely from a thermodynamic point of view, the Ruppeiner thermodynamic geometry is dealt with as a new attempt to extract the microscopic interaction information from the axioms of thermodynamics phenomenologically or qualitatively.

Considering the environment, i.e., an extensive, infinite thermodynamic system with entropy $S_e$, surrounding the black hole with entropy $S$, we can write the total entropy of the whole isolated system $S_{\text{total}}=S+S_e$. The $X^{\mu}$ correspond to some extensive variables of the black hole, like as mass $M$, charge $Q$, angular momentum $J$, and etc.. Correspondingly $X_e^{\mu}$ are extensities for the environment. $\Delta X^\mu\equiv X^{\mu}-X_0^{\mu}$ denotes the difference between $X^{\mu}$ and its equilibrium value $X_0^{\mu}$. We have known that in equilibrium state, the isolated thermodynamic system has a local maximum entropy. Hence for a small fluctuation $\Delta X^\mu$, $\Delta X_e^\mu$ away from this equilibrium, we have~\cite{Ruppeiner1995,Ruppeiner2007}
\begin{equation}\label{stot}
    \Delta S_{\text{total}}=\frac{\partial S}{\partial X^\mu}\Delta X^\mu
          +\frac{\partial S_e}{\partial X_e^\mu}\Delta X^\mu_e
      +\frac{1}{2}\frac{\partial^2 S}{\partial X^\mu \partial X^\nu}\Delta X^\mu \Delta X^\nu
       +\frac{1}{2}\frac{\partial^2 S_e}{\partial X_e^\mu \partial X_e^\nu}\Delta X^\mu_e \Delta X^\nu_e
       +\cdots.
\end{equation}
On the one hand, the conservation laws demand $\Delta X^\mu=-\Delta X_e^\mu$ and the maximum entropy needs a necessary condition $\partial S/\partial X^\mu=\partial S_e/\partial X_e^\mu$. On the other hand, if the environment is very large, i.e., $S \ll S_e \sim S_{\text{total}}$, the second quadratic term in Eq.~(\ref{stot}) is negligible compare with the first. Since the entropy $S_e$ of the environment as an extensive thermodynamical quantity gets the same order as that of the whole system, so its second derivatives with respect to the intensive thermodynamical quantities $x^\mu$ are much smaller than those of $S$ and has been ignored. Eventually we obtain an expression
\begin{equation}
\Delta S_{\text{total}}=-\frac{1}{2}\Delta l^2,
\end{equation}
where
\begin{equation}\label{line}
\Delta l^2=-\frac{\partial^2 S}{\partial X^\mu \partial X^\nu}\Delta X^\mu \Delta X^\nu.
\end{equation}

Looking at Eq.~(\ref{line}), the entropy as a thermodynamic potential can give rise to a thermodynamic line element very similar to the geometric one, which leads directly to a Ricci curvature scalar $R$. In this framework, there is an empirical observation that the thermodynamic curvature derived from the thermodynamic geometry~(\ref{line}) is related to the interaction of the system under consideration. More specifically, the negative (positive) thermodynamic curvature is associated with attractive (repulsive) microscopic interactions for an ordinary thermodynamic system. Though we know that the statement that the sign of the Ruppeiner curvature is related to the attractive or repulsive nature of the microscopic interaction is still a conjecture,  it is also verified for a large number of statistical physical models~\cite{Ruppeiner1995,Ruppeiner2014}.

For a black hole, because it can have temperature and entropy, this implies that a black hole has a microstructure. However, due to the lack of quantum gravity, there are certain assumptions in the study of microscopic properties of a black hole. From the point of view of thermodynamics to study the micro-behaviors of black holes, the current relevant researches\cite{Wei2015,Wei2019a,Wei2019b,Wei2019c,Miao2018a,Miao2018b,Miao2019a,Miao2019b,Aman2003,Mirza2007,Dehyadegari2017,Cai1999,Zhang2015a,
Zhang2015b,Liu2010,Xu2019a,Xu2019b,Niu2012,Wang2019,Ghosh2019,Bhattacharya2017,Chen2019,Guo2019,Mansoori2014,Mansoori2015,Mansoori2016,Sarkar2006,Quevedo2009,Wei2009,Akbar2011} show that the above thermodynamics geometry is a seemingly feasible scheme, which can be plausible to phenomenologically or qualitatively provide the information about interactions of black holes based on the well-established black hole thermodynamics. Hence we use this method to analyze the information of the charged BTZ black hole at present.

For black holes with AdS background, the most basic thermodynamic differential relation is $dM=TdS+VdP+\text{other work}$ and the term ``other work'' like as charge $Q$ and electrostatic potential $\Phi$, or angular momentum $J$ and angular velocity $\Omega$, or some coupling constants and their conjugations. Now we consider the situation with fixed other work terms. The form of the thermodynamic metric in Eq.~(\ref{line}) requires us to know $S=S(M,J,Q,...)$. Frequently, however, for the AdS black hole, we know instead $M=M(S,P,...)$. In this event, simplification results on writing the thermodynamic metric in the other thermodynamic potential, uaually with an additional pre-factor $1/T$. There are four conformally equivalent expressions of the above line element~(\ref{line}), which is particularly convenient for studying the thermodynamics geometry of AdS black holes. Before we start the process, let's make some comments.
\begin{itemize}
   \item The mass of black hole was initially regarded as the internal energy, but for the AdS black hole, due to the introduction of the extended phase space, i.e., thermodynamic pressure and volume, the mass of black hole corresponds to thermodynamic enthalpy. It's just a formal correspondence. It's ok to think of the enthalpy $M$ (or mass) as a conserved quantity. It is also extensive and additive.
  \item In thermodynamics, the extensive variable and the intensive variable are sometimes fuzzy. For example, if two ideal gas systems with the same pressure and volume are merged into a new system, which is the extensive variable for pressure and volume of the new system? If the pressure is kept constant during the merging, the volume of the total system multiplies and it is extensive and additive. Conversely if the volume is kept unchanged during the merging, the pressure of the total system multiplies and it is extensive and additive. Another very similar example is the voltage of the battery and the current in the circuit. If two batteries are connected in series, then the voltage multiplies and it is like the extensive variable; if two batteries are connected in parallel, then the current multiplies and it is like the extensive variable. The reason for this duality lies in the different ways of merging. In order to avoid this duality, we adopt the work of Quevedo~\cite{Quevedo2007}, that is, from a more mathematical (or geometric) language to define the extensive and intensive variables. It is considered that the quantities in differential form in the basic thermodynamic potentials are called extensive variables, and the corresponding conjugate quantities are intensive variables. Hence the extensive variable and the intensive variable are relative, and their roles can be exchanged for different thermodynamic potentials. For the internal energy $U$, if it is regarded as a basic thermodynamic potentials, we have $dU=TdS-PdV$ where the entropy $S$ and thermodynamic volume $V$ are extensive variables, while the temperature $T$ and thermodynamic pressure $P$ are intensive variables. If we take enthalpy $H$ as a basic thermodynamic potentials, we have $dH=TdS+VdP$ where the entropy $S$ and thermodynamic pressure $P$ are extensive variables, while the temperature $T$ and thermodynamic volume $V$ are intensive variables.
\end{itemize}

Hence for the AdS black hole, we can directly obtain the relation $dM=TdS+VdP$. It is the most basic thermodynamic differential relation of the AdS black hole. According to this expression, we have
\begin{eqnarray}
dS=\frac{1}{T}dM-\frac{V}{T}dP.
\end{eqnarray}
Hence under $S$ as a thermodynamic potential, $M$ and $P$ can be regarded as the extensive variables. We take $X^{\mu}=(M,P)$, and then the intensive variables corresponding to $X^{\mu}$ are $Y_{\mu}=\partial S/\partial X^{\mu}=(1/T,-V/T)$. It implies an isolated system of the black hole and environment with the same temperature and thermodynamic volume, where we have $M_{\text{total}}=M+M_e$ and $P_{\text{total}}=P+P_e$. This is very similar to the example of merging two ideal gas systems mentioned above. %If we require $P \ll P_e \sim P_{\text{total}}$, we can also obtain the condition $S \ll S_e \sim S_{\text{total}}$.}

Then the line element becomes
\begin{eqnarray}\label{xymetric}
\Delta l^2=-\frac{\partial^2 S}{\partial X^{\mu}\partial X^{\nu}}\Delta X^{\mu}\Delta X^{\nu}=-\Delta Y_{\mu} \Delta X^{\mu}.
\end{eqnarray}
Considering the specific form of each components in the above line element, we obtain
\begin{eqnarray}
\begin{aligned}
\Delta Y_0 &=\Delta\left(\frac{1}{T}\right)=-\frac{1}{T^2}\Delta T,\\
\Delta Y_1 &=\Delta\left(-\frac{V}{T}\right)=\frac{V}{T^2}\Delta T-\frac{1}{T}\Delta V.
\end{aligned}
\end{eqnarray}
Therefore, by inserting the above two expressions into Eq.~(\ref{xymetric}) and using the relation $\Delta M=T \Delta S+V \Delta P$, we finally write the line element as a universal form
\begin{eqnarray}\label{umetric}
\Delta l^2=\frac{1}{T}\Delta T \Delta S+\frac{1}{T}\Delta V \Delta P.
\end{eqnarray}

At present, we only consider the space composed of two generalized coordinates. According to the expression of the first law $dM=TdS+VdP$, we can see that there are four such coordinate spaces, which are $\{S, P\}$, $\{T, V\}$, $\{S, V\}$ and $\{T, P\}$. In the coordinate space $\{S,P\}$, we can see that
\begin{eqnarray}
\begin{aligned}
\Delta T &=\left(\frac{\partial T}{\partial S}\right)_P \Delta S+\left(\frac{\partial T}{\partial P}\right)_S \Delta P,\\
\Delta V &=\left(\frac{\partial V}{\partial S}\right)_P \Delta S+\left(\frac{\partial V}{\partial P}\right)_S \Delta P.
\end{aligned}
\end{eqnarray}
By introducing the above two expressions into the Eq.~(\ref{umetric}), we obtain \cite{Xu2019a}
\begin{eqnarray}\label{linesp}
\begin{aligned}
\Delta l^2 &=\frac{1}{T}\left(\frac{\partial T}{\partial S}\right)_P \Delta S^2+\frac{2}{T}\left(\frac{\partial T}{\partial P}\right)_S \Delta S \Delta P+\frac{1}{T}\left(\frac{\partial V}{\partial P}\right)_S \Delta P^2\\
&=g_{\mu\nu}\Delta x^{\mu}\Delta x^{\nu} \qquad (x^{\mu}=S,P),
\end{aligned}
\end{eqnarray}
where the Maxwell relation $(\partial T/\partial P)_{_S}=(\partial V/\partial S)_{_P}$ based on the relation $dM=TdS+VdP$ has been used.

For the line element in the coordinate space $\{S,V\}$, we need to use the differential relation of internal energy $U$, i.e., $dU=d(M-PV)=TdS-PdV$ and the Maxwell relation $(\partial T/\partial V)_{_S}=-(\partial P/\partial S)_{_V}$. Hence
\begin{eqnarray}\label{linesv}
\Delta l^2 =\frac{1}{T}\left(\frac{\partial T}{\partial S}\right)_V \Delta S^2+\frac{1}{T}\left(\frac{\partial P}{\partial V}\right)_S \Delta V^2=g_{\mu\nu}\Delta x^{\mu}\Delta x^{\nu} \qquad (x^{\mu}=S,V).
\end{eqnarray}
Similarly, in the coordinate space $\{T,V\}$, through the relation of the Helmholtz free energy $dF=-SdT-PdV$ and the Maxwell relation $(\partial S/\partial V)_{_T}=(\partial P/\partial T)_{_V}$, the line element is \cite{Xu2019a}
\begin{eqnarray}\label{linetv}
\begin{aligned}
\Delta l^2 &=\frac{1}{T}\left(\frac{\partial S}{\partial T}\right)_V \Delta T^2+\frac{2}{T}\left(\frac{\partial S}{\partial V}\right)_T \Delta T \Delta V+\frac{1}{T}\left(\frac{\partial P}{\partial V}\right)_T \Delta V^2\\
&=g_{\mu\nu}\Delta x^{\mu}\Delta x^{\nu} \qquad (x^{\mu}=T,V).
\end{aligned}
\end{eqnarray}
In the coordinate space $\{T,P\}$, through the relation of the Gibbs free energy $dG=-SdT+VdP$ and the Maxwell relation $(\partial S/\partial P)_{_T}=-(\partial V/\partial T)_{_P}$, the line element becomes
\begin{eqnarray}\label{linetp}
\Delta l^2=\frac{1}{T}\left(\frac{\partial S}{\partial T}\right)_P \Delta T^2+\frac{1}{T}\left(\frac{\partial V}{\partial P}\right)_T \Delta P^2=g_{\mu\nu}\Delta x^{\mu}\Delta x^{\nu} \qquad (x^{\mu}=T,P).
\end{eqnarray}

In terms of the metric $g_{\mu\nu}$ above mentioned, one can define the Christoffel symbols
\begin{eqnarray}
\Gamma^{\alpha}_{\beta\gamma}=\frac12 g^{\mu\alpha}
\left(\partial_{\gamma}g_{\mu\beta}+\partial_{\beta}g_{\mu\gamma}-\partial_{\mu}g_{\beta\gamma}\right),
\end{eqnarray}
and the Riemannian curvature tensors
\begin{eqnarray}
{R^{\alpha}}_{\beta\gamma\delta}=\partial_{\delta}\Gamma^{\alpha}_{\beta\gamma}-\partial_{\gamma}\Gamma^{\alpha}_{\beta\delta}+
\Gamma^{\mu}_{\beta\gamma}\Gamma^{\alpha}_{\mu\delta}-\Gamma^{\mu}_{\beta\delta}\Gamma^{\alpha}_{\mu\gamma}.
\end{eqnarray}Consequently the thermodynamic curvature, which is the ``thermodynamic analog'' of the geometric curvature in general relativity, is \begin{eqnarray}
R=g^{\mu\nu}{R^{\xi}}_{\mu\xi\nu}.
\end{eqnarray}

Obviously, the thermodynamic curvatures obtained by the above four line elements are equivalent to each other, because there is Legendre transformation between the thermodynamic potential functions corresponding to these different coordinate spaces. However, for the general form of thermodynamic curvature of the above four line elements respectively, the proof of equivalence of these thermodynamic curvatures seems to be more complicated. For simplicity, the equivalence can be illustrated by a specific example. Next, we will use the above four different coordinate spaces to calculate the thermodynamic curvature of the charged BTZ black hole respectively, so as to explore some micro-information that the black hole may have.

\section{Thermodynamic properties of charged BTZ black hole} \label{sec3}
For the $(2+1)$-dimensional charged BTZ black hole, its metric and the gauge field are \cite{Frassino2015,Mo2017}
\begin{eqnarray}
ds^2 &=&-f(r)dt^2+\frac{dr^2}{f(r)}+r^2 d\varphi^2, \nonumber\\
F&=&d A, \qquad A=-Q \ln\left(\frac{r}{l}\right)dt,
\end{eqnarray}
here the function $f(r)$ is
\begin{eqnarray}
f(r)=-2m-\frac{Q^2}{2}\ln\left(\frac{r}{l}\right)+\frac{r^2}{l^2}.
\end{eqnarray}
where $m$ is related to the black hole mass, $l$ is the AdS radius which is connected with the negative cosmological constant $\Lambda$ via $\Lambda=-1/l^2$ and $Q$ is the total charge of the black hole. About the basic thermodynamic properties in terms of the event horizon radius $r_h$ which is determined by the largest root of the equation $f(r_h)=0$, there are generally two different forms of these quantities.

\subsection{Case \Rmnum{1}}
By taking advantage of the Komar formula to determine the mass of the black hole, authors of Ref. \cite{Frassino2015} write the first law of thermodynamics of the charged BTZ black hole as $dM=TdS+VdP+\Phi dQ$, where the relevant quantities are
\begin{eqnarray}
M &=&\frac{m}{4}=\frac{r_h^2}{8l^2}-\frac{Q^2}{16}\ln\left(\frac{r_h}{l}\right),\label{enthalpy1}\\
T &=&\frac{r_h}{2\pi l^2}-\frac{Q^2}{8\pi r_h},\label{temperature}\\
S &=&\frac12 \pi r_h,\label{entropy}\\
P &=&-\frac{\Lambda}{8\pi}=\frac{1}{8\pi l^2}, \label{pressure}\\
V &=&\pi r_h^2-\frac14 \pi Q^2 l^2, \label{volume1}\\
\Phi &=&-\frac18 Q \ln\left(\frac{r_h}{l}\right).
\end{eqnarray}

\subsection{Case \Rmnum{2}}
An alternative scheme is renormalization procedure by enclosing the system in a circle of radius $r_0$ and taking the limit $r_0\rightarrow\infty $ whilst keeping the ratio $r/r_0=1$ \cite{Cadoni2008}. Then the black hole mass is interpreted as the total energy inside the circle of radius $r_0$. Based on this fact, Ref. \cite{Frassino2015} introduces a new thermodynamic parameter $R$ associated with the renormalization length scale $R=r_0$ via writing $$f(r)=-2m_0-\frac{Q^2}{2}\ln\left(\frac{r}{r_0}\right)+\frac{r^2}{l^2}.$$ The first law of thermodynamics of the charged BTZ black hole becomes $d\tilde{M}=TdS+\tilde{V}dP+\tilde{\Phi} dQ+K d R$, where the relevant quantities are
\begin{eqnarray}
\tilde{M} &=&\frac{m_0}{4}=\frac{r_h^2}{8l^2}-\frac{Q^2}{16}\ln\left(\frac{r_h}{R}\right),\label{enthalpy2}\\
\tilde{V} &=&\pi r_h^2, \label{volume2}\\
\tilde{\Phi} &=&-\frac18 Q \ln\left(\frac{r_h}{R}\right), \\
K &=&-\frac{Q^2}{16R},
\end{eqnarray}
and the other thermodynamic quantities $T$, $S$, $P$ are still expressions~(\ref{temperature}),~(\ref{entropy}) and~(\ref{pressure}).

\section{Thermodynamic curvature of charged BTZ black hole} \label{sec4}
In a thermodynamic system, the phase space contains both the generalized coordinates and their conjugate generalized forces (or the conjugated extensive and intensive quantities). Depending on the selection of the thermodynamic potential, the role of the two can be exchanged. Here the phase space of the BTZ black hole should be $\{T, S, P, V, \Psi, Q\}$ in Case \Rmnum{1} and $\{T, S, P, \tilde{V}, \tilde{\Phi}, Q, K, R\}$ in Case \Rmnum{2}. Once the thermodynamic potential is chosen, when a certain generalized coordinates are fixed, naturally, the conjugate generalized force can be obtained and also fixed directly.

For our present discussion, we calculate the thermodynamic curvature in two-dimensional coordinate space, in which the conjugate generalized forces are naturally emerged. On the one hand, as an analogy with a simple thermodynamic system, the influence of pressure-volume and temperature-entropy on the thermodynamic behavior of the system seems to be paid more attention. On the other hand, it is found that the results we obtained for the fixed charge and for the charge as an independent thermodynamic quantity are qualitatively consistent. Therefore, in order to simplify the calculation without losing some physical information, we limit the analysis to fixed charge $Q$ in Case \Rmnum{1} and fixed $Q$ and $R$ in Case \Rmnum{2}.

In addition, as supplementary material, we also show the effect of the charge $Q$ on the behavior of the thermodynamic curvature of the charged BTZ black hole in Appendix \ref{app}. The qualitative behavior of thermodynamic curvature of the charged BTZ black hole is consistent whether in the coordinate space $\{S,P\}$ with fixed charge $Q$ or in the coordinate spaces $\{S,Q\}$ and $\{S, P, Q\}$ with the charge $Q$ as an independent thermodynamic quantity. We all have positive thermodynamic curvature.

\subsection{Thermodynamic curvature for Case \Rmnum{1}}
In principle, there should be four ways, i.e., in coordinate spaces $\{S,P\}$, $\{S,V\}$, $\{T,V\}$ and $\{T,P\}$. For the coordinate space $\{T,V\}$, we need to write the entropy $S$ and thermodynamic pressure $P$ as a function of temperature $T$ and thermodynamic volume $V$, respectively. The analytical forms become very complicated, hence we will ignore the case here only for avoiding the technique complexity.
\begin{itemize}
  \item In the coordinate space $\{S,P\}$, we need to write the temperature $T$ and thermodynamic volume $V$ as functions of entropy $S$ and pressure $P$, respectively
      \begin{eqnarray}
      T=\frac{128PS^2-\pi Q^2}{16\pi S}, \qquad V=\frac{128PS^2-\pi Q^2}{32\pi P}.
      \end{eqnarray}
      Hence according to Eq.~(\ref{linesp}), we can directly calculate the expression of thermodynamic curvature
      \begin{eqnarray}
      R_{SP}=\frac{384\pi Q^2 PS}{(\pi Q^2+256P S^2)^2}.
      \end{eqnarray}
  \item In the coordinate space $\{S,V\}$, the temperature $T$ and thermodynamic pressure $P$ can be written as functions of entropy $S$ and volume $V$, respectively
      \begin{eqnarray}
      T=\frac{\pi Q^2 V}{16S(4S^2-\pi V)}, \qquad P=\frac{\pi Q^2}{128 S^2-32\pi V}.
      \end{eqnarray}
      Hence according to Eq.~(\ref{linesv}), we can obtain the expression of thermodynamic curvature
      \begin{eqnarray}
      R_{SV}=\frac{12S(4S^2-\pi V)}{(\pi V-12 S^2)^2}.
      \end{eqnarray}
  \item In the coordinate space $\{T,P\}$, we must have the expressions of entropy $S$ and thermodynamic volume $V$ in terms of temperature $T$ and pressure $P$, respectively
      \begin{eqnarray}
      S=\frac{\pi T+\sqrt{\pi^2 T^2+2\pi Q^2 P}}{16P}, \qquad V=\frac{T}{32P^2}\left(\pi T+\sqrt{\pi^2 T^2+2\pi Q^2 P}\right).
      \end{eqnarray}
      Hence according to Eq.~(\ref{linetp}), we can obtain the expression of thermodynamic curvature
      \begin{eqnarray}
      R_{TP}=\frac{24P\left[8\pi T^2\left(-\sqrt{\pi}T+\sqrt{2Q^2 P+\pi T^2}\right)+3Q^2 P\left(-5\sqrt{\pi}T+3\sqrt{2Q^2 P+\pi T^2}\right)\right]}{\sqrt{\pi}(9Q^2 P+4\pi T^2)^2}.
      \end{eqnarray}
\end{itemize}

Meanwhile, we can easily test the identity as desired
\begin{eqnarray}
R_{SP}=R_{SV}=R_{TP}>0.
\end{eqnarray}
Furthermore, for the extreme black hole $T=0$, we can observe clearly that thermodynamic curvature is a finite positive value $R_{_{T=0}}=1/(3S)$. Based on an empirical conclusion under the framework of thermodynamic geometry theory, i.e., the negative (positive) thermodynamic curvature is associated with attractive (repulsive) microscopic interactions for a thermodynamic system \cite{Ruppeiner2008,Wei2015,Wei2019a,Wei2019b,Wei2019c,Miao2018a,Miao2018b,Miao2019a,Miao2019b,Xu2019a,Xu2019b}, we can speculate that the charged BTZ black hole is likely to present a repulsive between its molecules phenomenologically or qualitatively. In addition, when $Q=0$, the thermodynamic curvature degenerates to zero, which implies that the neutral BTZ black hole is Ruppeiner flat \cite{Cai1999,Sarkar2006,Akbar2011}.

\subsection{Thermodynamic curvature for Case \Rmnum{2}}
Generally, there should also be four ways, i.e., in coordinate spaces $\{S,P\}$, $\{T,\tilde{V}\}$, $\{S,\tilde{V}\}$ and $\{T,P\}$. However, because entropy $S$ and thermodynamic volume $\tilde{V}$ are not independent of each other, the coordinate space $\{S,\tilde{V}\}$ is invalid.
\begin{itemize}
  \item In the coordinate space $\{S,P\}$, we need to write the temperature $T$ and thermodynamic volume $\tilde{V}$ as functions of entropy $S$ and pressure $P$, respectively
      \begin{eqnarray}
      T=\frac{128PS^2-\pi Q^2}{16\pi S}, \qquad \tilde{V}=\frac{4S^2}{\pi}.
      \end{eqnarray}
      Hence according to Eq.~(\ref{linesp}), we can directly calculate the expression of thermodynamic curvature
      \begin{eqnarray}
      \tilde{R}_{SP}=\frac{2\pi Q^2}{S(128PS^2-\pi Q^2)}.
      \end{eqnarray}
  \item In the coordinate space $\{T,\tilde{V}\}$, the entropy $S$ and thermodynamic pressure $P$ can be written as functions of temperature $T$ and volume $\tilde{V}$, respectively
      \begin{eqnarray}
      S=\left(\frac{\pi \tilde{V}}{4}\right)^{1/2}, \qquad P=\frac{\sqrt{\pi}T}{4\sqrt{\tilde{V}}}+\frac{Q^2}{32\tilde{V}}.
      \end{eqnarray}
      Hence according to Eq.~(\ref{linetv}), we can obtain the expression of thermodynamic curvature
      \begin{eqnarray}
      \tilde{R}_{T\tilde{V}}=\frac{Q^2}{2\pi T\tilde{V}}.
      \end{eqnarray}
  \item In the coordinate space $\{T,P\}$, we must have the expressions of entropy $S$ and thermodynamic volume $\tilde{V}$ in terms of temperature $T$ and pressure $P$, respectively
      \begin{eqnarray}
      S=\frac{\pi T+\sqrt{\pi^2 T^2+2\pi Q^2 P}}{16P}, \qquad \tilde{V}=\frac{\pi T^2+Q^2 P+T\sqrt{\pi^2 T^2+2\pi Q^2 P}}{32P^2}.
      \end{eqnarray}
      Hence according to Eq.~(\ref{linetp}), we can obtain the expression of thermodynamic curvature
      \begin{eqnarray}
      \tilde{R}_{TP}=\frac{16 Q^2 P+16T\left(\pi T-\sqrt{\pi^2 T^2+2\pi Q^2 P}\right)}{\pi Q^2 T}
      \end{eqnarray}
\end{itemize}
Meanwhile we also have the relation as desired
\begin{eqnarray}
\tilde{R}_{SP}=\tilde{R}_{T\tilde{V}}=\tilde{R}_{TP}>0.
\end{eqnarray}
Moreover, for the extreme black hole $T=0$, we can observe clearly that thermodynamic curvature tends to be positive infinity. Hence we can conjecture that the charged BTZ black hole is likely to present a repulsive between its molecules phenomenologically or qualitatively. When $Q=0$, the thermodynamic curvature vanishes, indicating that the neutral BTZ black hole is Ruppeiner flat \cite{Cai1999,Sarkar2006,Akbar2011}. More importantly, comparing with the thermodynamic curvature of Case \Rmnum{1} in which the thermodynamic curvature of extreme black hole tends to be a finite positive value, we can clearly observe that this difference can be used as a diagnosis for the discrimination of two different thermodynamic approaches of the charged BTZ black hole.

\section{Conclusion and Discussion}\label{sec5}
Comparing the results of Case \Rmnum{1} and Case \Rmnum{2}, we can conclude that:
\begin{itemize}
  \item Both of Case \Rmnum{1} and Case \Rmnum{2}, the resulting thermodynamic curvatures are always positive, which may be related to the information of repulsive interaction between black hole molecules for the charged BTZ black hole.
  \item When $Q=0$, in both cases, we can see that the thermodynamic curvature degenerates to zero, which shows that the neutral BTZ black hole is Ruppeiner flat \cite{Cai1999,Sarkar2006,Akbar2011}.
  \item For the extreme black hole, i.e., $T=0$, the thermodynamic curvature is a finite positive value in Case \Rmnum{1}, while in Case \Rmnum{2}, it tends to be positive infinity. In the previous analysis of the thermodynamic curvature of AdS black holes \cite{Wei2015,Wei2019a,Wei2019b,Wei2019c,Miao2018a,Miao2018b,Miao2019a,Miao2019b,Xu2019a,Ghosh2019}, we usually see that in extreme black holes, the thermodynamic curvature tends to be positive or negative  infinity. In Case \Rmnum{1} of the thermodynamic analysis about charged BTZ black hole, the thermodynamic curvature of the extreme black hole is a positive finite value, which is not consistent with the previous discussion. From this point of view, thermodynamic curvature of the extreme black hole discussed in the present paper may serve as a criterion to discriminate the two thermodynamic approaches introduced in Ref. \cite{Frassino2015} and our result seems to support the Case \Rmnum{2} of the thermodynamic analysis about charged BTZ black hole. This is consistent with the result obtained in \cite{Mo2017} by using the idea of holographic heat engine.
\end{itemize}

We have known that for Case \Rmnum{1}, the charged BTZ solution is proved to be a super-entropic black hole \cite{Johnson2019a,Johnson2019b}, which violates the reverse isoperimetric inequality. Generally, for a $d$-dimensional black hole, its thermodynamic volume $V$ and entropy $S$ satisfy the reverse isoperimetric inequality \cite{Cvetic2011}
$$\mathcal{R}=\left(\frac{(d-1)V}{\omega_{d-2}}\right)^{\frac{1}{d-1}}\left(\frac{\omega_{d-2}}{4S}\right)^{\frac{1}{d-2}}\geq 1,$$ where $\omega_n=2\pi^{(n+1)/2}/\Gamma\left[(n+1)/2\right]$ is the standard volume of the round unit sphere. In $d=3$, the charged BTZ black hole has $\mathcal{R}<1$ in Case \Rmnum{1}. While in Case \Rmnum{2}, because the thermodynamic volume $\tilde{V}$ and entropy $S$ are not independent of each other, then the above reverse isoperimetric inequality is saturated by the charged BTZ black hole, i.e., $\mathcal{R}=1$. The underlying reason for the satisfaction, violation or saturation of the reverse isoperimetric inequality lies in the definition of the thermodynamic volume of a black hole, which is also a very significant research issue in black hole thermodynamics. In general, if the thermodynamic volume is consistent with the expression of the geometric volume, often the reverse isoperimetric inequality is saturated ($\mathcal{R}=1$), while when the thermodynamic volume of the black hole does not look like any geometric volume, it generally corresponds to a super-entropic black hole ($\mathcal{R}<1$) or sub-entropic black hole ($\mathcal{R}>1$).

When {\em the reverse isoperimetric inequality is saturated, the thermodynamic curvature of extreme black hole tends to be (positive or negative) infinity}. This conjecture is verified by many examples, like Schwarzschild AdS black hole, Reissner-Nordstr\"{o}m AdS black hole and Gauss-Bonnet AdS black hole (and various other simple static black hole solutions of the pure Einstein gravity or higher-derivative generalizations thereof), and Case \Rmnum{2} of our current discussing. Moreover, according to the above calculation and analysis of Case \Rmnum{1} for the charged BTZ black hole,
these results are likely to hint an empirical conjecture that {\em for super-entropic black holes, the thermodynamic curvature of extreme black hole will go to a finite (positive or negative) value.} In the future work, we will test the conjecture in other super-entropy black holes, like ultra-spinning limit of Kerr-AdS black holes \cite{Hennigar2015a,Hennigar2015b}. Meanwhile, for the sub-entropic black hole, such as the Kerr-AdS black hole \cite{Cvetic2011,Johnson2019c}, STU black holes \cite{Johnson2019c,Caceres2015}, Taub-NUT/Bolt black hole \cite{Johnson2014}, generalized exotic BTZ black hole \cite{Johnson2019b}, noncommutative black hole \cite{Miao2017} and accelerating black holes\cite{Appels2016}, what kind of conjecture can we make? These are also very interesting topics to be discussed in the future.

\section*{Acknowledgments}
The financial supports from National Natural Science Foundation of China (Grant Nos.11947208 and 11947301), China Postdoctoral Science Foundation (Grant No. 2020M673460), Major Basic Research Program of Natural Science of Shaanxi Province (Grant No.2017ZDJC-32), Scientific Research Program Funded by Shaanxi Provincial Education Department (Program No.18JK0771) are gratefully acknowledged. This research is supported by The Double First-class University Construction Project of Northwest University. The authors would like to thank the anonymous reviewers for the helpful comments that indeed greatly improve this work.

\appendix
\section{Thermodynamic curvature of the charged BTZ black hole in the coordinate spaces $\{S,Q\}$ and $\{S, P, Q\}$}\label{app}
In order to make the analysis and discussion of this paper more complete, we now consider the situation of the charge $Q$ as an independent thermodynamic variable. Here we just limit the calculation to the Case \Rmnum{1}. For the Case \Rmnum{2}, there will be similar results.

In the $\{S,Q\}$ space with fixed the AdS radius $l$, we obtain the line element of thermodynamic geometry
\begin{eqnarray}
\begin{aligned}
\Delta l^2 &=\frac{1}{T}\left(\frac{\partial T}{\partial S}\right)_Q \Delta S^2+\frac{2}{T}\left(\frac{\partial T}{\partial Q}\right)_S \Delta S \Delta Q+\frac{1}{T}\left(\frac{\partial \Phi}{\partial Q}\right)_S \Delta Q^2\\
&=g_{\mu\nu}\Delta x^{\mu}\Delta x^{\nu} \qquad (x^{\mu}=S,Q).
\end{aligned}
\end{eqnarray}
In order to keep the generality and make the calculation simple and clear, we set $l=1/\pi$. Hence the thermodynamic curvature reads as
\begin{eqnarray}
R_{SQ}=\frac{A+B\ln (2 S)+C\ln^2 (2 S)}{2 S\left(16 S^2-Q^2\right) \left[\left(Q^2+16 S^2\right) \ln (2 S)+2 Q^2\right]^2},
\end{eqnarray}
where
\begin{eqnarray*}
\begin{aligned}
A=&\left(3 Q^2+16 S^2\right) \left(Q^4+64 Q^2 S^2-256 S^4\right),\\
B=&Q^6+336 Q^4 S^2+2816 Q^2 S^4-4096 S^6,\\
C=&32\left(3 Q^4 S^2+64 Q^2 S^4+256 S^6\right).
\end{aligned}
\end{eqnarray*}
After a little discussion, it can be seen that the thermodynamic curvature $R_{SQ}$ is always positive under the condition $T>0$, i.e., $4S>Q$. This means that the charged BTZ black hole is still dominated by repulsion in the $\{S,Q\}$ space.

Next, we extend the research to three-dimensional space. In the $\{S,P,Q\}$ space, the metric $g_{\mu\nu}$ of thermodynamic geometry is
\begin{eqnarray}
\begin{aligned}
g_{\mu\nu}=\frac{1}{T}\left(
\begin{array}{ccc}
 \frac{\partial T}{\partial S} & \frac{\partial T}{\partial P} & \frac{\partial T}{\partial Q} \\
 \frac{\partial V}{\partial S} & \frac{\partial V}{\partial P} & \frac{\partial V}{\partial Q} \\
 \frac{\partial \Phi}{\partial S} & \frac{\partial \Phi}{\partial P} & \frac{\partial \Phi}{\partial Q} \\
\end{array}
\right).
\end{aligned}
\end{eqnarray}
Accordingly, we can calculate the thermodynamic curvature directly
\begin{eqnarray}
R_{SPQ}=\frac{D+E\ln \left(\frac{32 P S^2}{\pi }\right)+F\ln ^2\left(\frac{32 P S^2}{\pi }\right)}{2 S\left(128 P S^2-\pi  Q^2\right) \left[\left(256 P S^2+\pi  Q^2\right) \ln \left(\frac{32 P S^2}{\pi }\right)+6 \pi  Q^2\right]^2},
\end{eqnarray}
where
\begin{eqnarray*}
\begin{aligned}
D=&6 \left(-4194304 P^3 S^6+81920 \pi  P^2 Q^2 S^4+1536 \pi ^2 P Q^4 S^2+9 \pi ^3 Q^6\right),\\
E=&4 \left(-4194304 P^3 S^6+286720 \pi  P^2 Q^2 S^4+2240 \pi ^2 P Q^4 S^2+3 \pi ^3 Q^6\right),\\
F=&8388608 P^3 S^6+425984 \pi  P^2 Q^2 S^4+640 \pi ^2 P Q^4 S^2+\pi ^3 Q^6.
\end{aligned}
\end{eqnarray*}
Similarly, the obtained thermodynamic curvature $R_{SPQ}$ is also positive under the condition $T>0$, i.e., $128 P S^2>\pi  Q^2$, which means that the charged BTZ black hole is related to the information of repulsive interaction between black hole molecules in the $\{S,P,Q\}$ space.

\end{document}